# Identifying Security Risks in NFT Platforms

Yash Gupta and Jayanth Kumar

# Abstract

**Purpose**: This paper examines the effects of inherent risks in the emerging technology of non-fungible tokens and proposes an actionable set of solutions for stakeholders in this ecosystem and observers. Web3 and NFTs are a fast-growing 300 billion dollar economy with some clear, highly publicized harms that came to light recently. We set out to explore the risks to understand their nature and scope, and if we could find ways to mitigate them.

**Method**: In due course of investigation, we recap the background of the evolution of the web from a client-server model to the rise of Web2.0 tech giants in the early 2000s. We contrast how the Web3 movement is trying to re-establish the independent style of the early web. In our research we discover a primary set of risks and harms relevant to the ecosystem, and classify them into a simple taxonomy while addressing their mitigations with solutions.

**Results:** We arrive at a set of solutions that are a combination of processes to be adopted, and technological changes or improvements to be incorporated into the ecosystem, to implement risk mitigations. By linking mitigations to individual risks, we are confident our recommendations will improve the security maturity of the growing Web3 ecosystem.

**Implications**: The Web 3 and NFT movement isn't just about trading digital goods, but is creating fundamental new capabilities for our Internet, helping people establish clear ownership of digital goods. Establishing security best practices will enable serious minimization of harms in this nascent technology.

**Disclaimer**: We are not endorsing, or recommending specifically any particular product or service in our solution set. Nor are we compensated or influenced in any way by these companies to list these products in our research. The evaluations of products in our research have to simply be viewed as suggested improvements.

## Background

Multiple factors have contributed to the rise of alternative investments in recent years instead of traditional options (Rastegar, 2021). For this paper, we consider traditional investments to be: stocks, bonds, derivatives; whereas alternative investments cover anything else. The cryptocurrency market is a new and well-known example of an alternative investment. The inherent value of Bitcoin is a significant factor that influences the entire crypto market. As of December 2021, Bitcoin has increased by almost 9000% since 2016. This development is due to various factors including the adoption of cryptocurrency by various government bodies and commercial firms throughout the World. With the rise of Bitcoin and the crypto market, a new type of investment has emerged: Non-Fungible Tokens (NFTs).

NFTs are inherently difficult to comprehend. We will start by looking at what non-fungible means. The term "non-fungible" refers to something distinctly unique from others and cannot be altered. NFTs are a type of cryptocurrency that is part of a particular blockchain, and the bulk of NFTs today currently function on top of the Ethereum blockchain, which means they are traded in Ethereum. A cryptocurrency differs from an NFT in that the NFT can store considerably more data, such as digital art or music. Anyone can develop NFTs and trade them, and anything can be an NFT. To trade NFTs, register an account on the NFT markets portal and fill in all the details to create a wallet (Clark, 2021). Leading marketplaces such as OpenSea and Nifty Gateway have made the possibility widely available to the general public, greatly simplifying the procedure.

OpenSea and Nifty Gateway were among the first to leverage NFTs by establishing popular marketplaces for them, giving them an advantage against new platforms that trade in various cryptocurrencies. Many other platforms attempted to popularize NFT clients, such as NBA TopShots with Dapper Labs, who created NFTs that caught NBA events to diversify the customer base but struggled (Sarlin, 2021). However, things have recently changed, with major corporations such as Disney focusing on the NFT sector. Many OpenSea and Nifty Gateway customers might switch to Disney and Veve. Furthermore, with Facebook orienting its company around virtual reality and Nike entering the digital footwear market, there will undoubtedly be a significant impact on OpenSea (Bloomberg, 2021).

Platforms like OpenSea are also impacted by the entry of prominent players in this market and the security threats they confront due to possible weaknesses in their systems. OpenSea has recently been in the headlines due to security vulnerabilities that might have resulted in significant losses for its consumers. A problem in OpenSea was recently discovered, allowing attackers to steal a customer's entire crypto wallet simply by sending a malicious NFT. This vulnerability would result in many pop-ups from OpenSea's storage domain. Because the URL was fraudulent, the attacker could access the victim's full crypto wallet and steal all their funds (Dent, 2021).

This research paper explores closely the types of malicious attacks that are ultimately not dependent on an inherent flaw in the underlying system or technology implementation. These risks manifest because they are dependent on the human interacting with the system or technology. Our paper accumulates them under a standard banner designated as 'Layer-8 risks'.

Although it is unknown whether or not the vulnerability discussed earlier was exploited, the fact that it was discovered compelled OpenSea to fix it and assess the impact on business. They also issued a statement in which they stated that they would increase their efforts to educate their employees about the importance of security. NFT platforms lack good security practices, making them unable to accurately reflect their company's risk appetite and preventing them from developing an effective control architecture to safeguard against external and internal threats.

## Historical Background

The concerns regarding digital ownership of goods are not new, which is why one of the core security concerns attempting to be addressed within the NFT market space is the proper assignment, protection and validation of ownership. The Internet has a long history of struggling to protect digital asset owners' rights adequately. To understand the new gold rush around trading in NFT's we need to recall that since its inception, the Internet did not provide a reliable way to establish ownership of digital goods such as music and artwork. Any rights management systems such as DRM (Digital Rights Management) that prevented unauthorized copying and sharing were bolted-on to the applications that host or transmit the content.

On the Internet, digital goods are treated as pure information. To ensure that information is easily transportable and interpretable by a wide range of machines across the World, it has to be easily shared and stored in multiple formats. Easy reproducibility also introduces a problem of unauthorized copying, and no mechanism to prevent or differentiate between authorized sales and unauthorized ownership.

When content sharing became popular, the music industry wanted to 'patch' the Internet with Digital Rights Management solutions to ensure that piracy and casual content sharing do not eat into conventional content sales like compact discs (CDs) and cassettes.

In the late 90s, there was an inherent distrust between content creators and distributors like film studios, and record labels, giving rise to the use and adoption of litigation, famously by Metallica against Napster (Reed, 2018).

The result of this lawsuit paved the way for record labels to insist on the usage of DRM technology to limit unauthorized copy and use of music purchased from stores like iTunes. DRM was quite unpopular among users and music lovers. The technology itself had fundamental contradictions in how it was applied. So much so that Steve Jobs, former CEO of Apple, famously outlined in his' Thoughts on Music' memo, "*Though the big four music companies require that all their music sold online be protected with DRMs, these same music companies continue to sell billions of CDs (compact disc) a year which contain completely unprotected music. That's right! No DRM system was ever developed for the CD, so all the music distributed on CDs can be easily uploaded to the Internet, then (illegally) downloaded and played on any computer or player.*" (Jobs, 2007).

DRM rapidly fell out of popularity because it limited users while also offering minimal protections against illegal usage of said content. The unpopularity of DRM was so widespread that while it was still in use in subtle forms across digital content, there were no significant improvements to the model or research for establishing ownership verification on the Internet after the mid-2000s, until the early 2020's.

## Background Summary

As discussed earlier, NFTs were first made possible by innovations over the blockchain, specifically Ethereum technology. Ethereum is a community-built technology behind the cryptocurrency ether (ETH). Anyone can send cryptocurrency to someone else for a small service fee using this technology. Since Ethereum is programmable, it is used for multiple innovative applications.

If adequately implemented, NFTs and NFT marketplaces could solve many problems associated with ownership records on the Internet by maintaining a public record of the ownership of the content of the NFT on the blockchain.

Content creators can also directly sell their content on a global market. The availability of ownership verification enables the usage of digital goods, just like you would use physical-world goods. You can place a digital NFT as collateral for a loan, just like you would with an expensive work of art.

Web2 is the conventional Internet that enables computers to connect and communicate across the World. In the conventional model of the World Wide Web, networks of computers were interconnected based on their usage of the TCP/IP protocol to communicate commonly using a client-server architecture. The strengths of the TCP/IP protocol enable the successful communication of billions of devices at scale due to the primary strengths of the protocol. These are:

1. Integrated addressing system enables identification and addressing of systems irrespective of how the underlying network is constructed (Kozierok, 2005)
2. TCP/IP is designed for the routing of data over a network of arbitrary complexity (Kozierok, 2005)
3. TCP/IP is independent of the underlying network and is the most popular protocol for connecting networks and devices due to its high scalability

In the second iteration of the world wide web, Web 2.0, the network inherently became more centralized due to the rise of private platforms like Google, Amazon, Facebook, and others. Since private firms are driven by profit, they focus on the client/server model. In the late 20th century, it was thought that the separation of capabilities of a few powerful 'server' computers and many distributed client PCs would be ideal for delivering the multimedia experiences that constituted the Internet at a scale that reached a global audience.

By distributing the complexity among a few powerful servers, and relatively inexpensive client machines, the requirements on an individual user or node were minimal. With maintenance, development, and processing primarily being performed on the server-side, websites and services started growing drastically on the web.

In the early 2000s, the Internet was still an inherently decentralized landscape, with several exciting technologies like weblogs (Blogs), personal web pages, and small businesses. However, with the rise of private players like Google, Facebook, and Yahoo (at the time) rapidly innovating, most users congregated to the 'tech giants,' resulting in the slow decline of what was once a decentralized network of users. The Internet in its present form is only a few platforms with billions of users that are essentially walled gardens. Web3 proponents state that they are trying to go back to the essence of what the Internet was supposed to be, a decentralized ecosystem empowering individuals.

There are many different definitions for Web3, and there is still some debate. Web3 refers to a decentralized ecosystem of technologies and applications based on the blockchain (Edelman, 2021). The main critique that is posed to Web 2 is the 'centralization' of the Internet at the hands of proprietary companies ("Web2 v Web3," 2021). Web3 apps, known in the Ethereum domain as dApps or decentralized apps, are not centrally administered, and thus, they offer more liberties to the participants on their intended use. There can be no gatekeepers to the access and usage of decentralized apps, which is a significant evolution over Web 2.0.

# Problem Statement

The rise of decentralized technologies like blockchain has resulted in a $370 billion economy purely for creating and trading non-fungible tokens (NFTs). Platforms and participants in this new economy are becoming high-profile victims of scams and bugs while having no clear way to redress financial grievances. A significant portion of the losses within this ecosystem is due to NFT market leaders' delay in implementing currently recognized security best practices that would adequately protect user ownership of digital assets. Furthermore, the mix of centralized and decentralized technology creates novel security concerns within the NFT marketplace that leaders have been slow to address.

# Research on Problem

The core of cybersecurity is protecting assets' confidentiality, integrity, and availability. These principles apply equally to the innovations on Web3 technologies like NFTs. A few requirements for applying measurable value and ownership to digital goods are properly implemented ownership, authentication, and authorization. Additionally, the integrity of the ownership record and the asset itself is critical in developing a thriving digital goods economy on these paradigms. Our initial research indicates that the current technological landscape, along with certain technologies addressing the security concerns of NFTs, can be leveraged to address the issues of storage and ownership of NFTs while also ensuring the security of these NFT platforms.

The client/server model provided some critical advantages to the affordability of accessing the World Wide Web when computing resources came at a high cost. In the current state of the web, security issues stem from the centralization that inherently arises from the client-server model. In a network with one powerful central server and client machines, the processing power, maintenance, and upkeep can be restricted to a few critical machines. This arrangement allows the client machines to be relatively low-powered and even allows computers with low bandwidth to access the Internet. Ultimately, network centralization was created to improve efficiency and take advantage of potential economies of scale (N-Able, 2018).

With the rise in the walled garden techno-landscapes from the mid-2000s, such as YouTube, Facebook, and others, we now have a digital ecosystem shared on a centralized Web 2.0 platform. Typically, user terms and conditions state that users remain the owner of the content they share by agreeing to the terms. In effect, users grant the platform a "non-exclusive, royalty-free, transferrable, sub-licensable, worldwide license" to use any shared content (Blakers, 2017). The platform may monetize it, and there is nothing that the creator can do about this unless there are remedial steps such as filing a Digital Millennium Copyright Act (DMCA) complaint or pursuing legal action. In nations where copyright enforcement is not reliable or prompt, these options may also be not worth pursuing at scale.

Reliable storage is another concerning issue to be addressed with digital content. Content on the web is stored either in physical servers or cloud infrastructures in the client/server model, making the content retrievable but prone to data loss issues in cases where data redundancy is not maintained. This kind of centralized storage currently implemented by developers of NFTs hosted on OpenSea usually has many weaknesses. However, the two that affect the security and this research area are impermanence and mutability.

Mutability is a critical aspect of the NFT's metadata. If the metadata is not frozen, the developer can modify the image to something that the collector is unaware of or can lead to loss of financial value of the NFT (Atallah, 2021). For example, suppose the cloud is owned by a single entity, i.e., a cloud provider. In that case, the content can disappear when the server or cloud service goes offline, resulting in the collector no longer having access to the content. Additionally, these storage options are always vulnerable to ransomware attacks where malicious actors deliberately target high-value information databases to extort ransoms.

## Ensuring Integrity and Availability of Digital Goods:

NFTs have faced issues verifying ownership and storage of unique items in the past. However, with Ethereum's ERC-721 (Entriken, 2018), the verification issues of ownership have been remedied with a single token ID that can verify different attributes of a unique NFT. The issue of safe and reliable storage still poses many threats which cause the consumers to lose money.

Most of the NFTs are minted on Ethereum, so the NFT is represented by an indivisible, immutable, and unique address on the Ethereum network. However, the content and metadata of the NFT are not represented in it; they still exist on the permanent storage network. More precisely, the NFT lives on the blockchain, whereas the content lives on a centralized server which puts the NFTs to face the same issue as content on Web 2.

## A Taxonomy of Risks in the NFT Landscape:

Owners of NFTs and the prominent platforms that enable NFT commerce are the targets that we shall focus our research efforts on. In order to understand the behavior of malicious actors, we will first propose a preliminary classification of risks in the NFT space based on the most popular methods of conducting attacks on NFT owners, platforms, and other entities in the ecosystem. We can learn from existing standards of classifying operational cybersecurity risk and apply that to the NFT platforms. We believe that malicious attacks on NFTs can be classified on this basis given below:

| **Layer 8 Risks** | **Missing or Failing Processes** | **External Risks** |
|---|---|---|
| 1.1 **Inadvertent** : User error related risks<br><br>1.2 **Malicious attacks** : Phishing scams or purchasing bots<br><br>1.3 **Scams** | 2.1 **Process Controls** : Quality control processes such as security review<br><br>2.2 **Partnerships**<br><br>2.3 **Platform bugs** | 3.1 **Regulatory risk**<br><br>3.2 **Disasters** resulting in partial/total loss of content or ownership record<br><br>3.3 **Geopolitical risk** |

While we developed this taxonomy of risks, we referenced a taxonomy of operational risks developed by the Software Engineering Institute at Carnegie Mellon University to classify risks in the NFT space.

## Layer-8 Risks:

Currently, there is no Layer-8 in the conventional OSI model of computer networks; however, some intelligent and amusing engineers once coined the term to mean anything related to the user that is not included in the seven layers of the OSI model. We shall use the term Layer-8 risks to signify user-level errors that cannot be described as part of the conventional architecture.

It is estimated that historically, a tiny percentage of security controls contained points about educating and training users and people that are inherent in systems and processes. An excerpt from 'CIO report' states that the number may be less than 5% of total security controls (di Fabio, n.d.). It is, therefore, not a big surprise that conventional software development lifecycle processes (SDLC) inherently introduced a significant number of unforeseen vulnerabilities that may result in the exploitation of end-users and result in significant harm and loss.

Additionally, modern threats and threat actors are unlike the early threats people encountered in the late 90s/early 2000s. Today's threat actors act and move more like a real human adversary would in a battle engagement, performing reconnaissance on targets, choosing the techniques, tools, and procedures of engagement uniquely for different kinds of victims.

The MITRe ATT&CK framework is a well-renowned framework on attack procedures that documents 20+ techniques that modern adversaries conduct prior to even intrusion, to understand the profile of their targeted victim. The Verizon 2021 Data Breach report also states that a significant portion of detected intrusions in breaches occurred due to phishing attacks or due to lost or stolen credentials, indicating that the use of malware in intrusions itself is fast becoming a thing of the past. Of course, they are still used once the adversary has infiltrated the target system or platform.

Layer-8 risks are a genuine consideration for the Web3 and decentralization movement, especially for OpenSea and other such NFT platforms. Increasingly sophisticated threat actors that target conventional systems are now applying their time and energies towards exploiting the user-base of said platforms. As mentioned earlier, the famous example of a phishing loophole in the pre-eminent NFT platform, OpenSea, is an excellent example of a blend of both of our risks. We outline it in detail in the next section.

**1.1 Inadvertent or User Error Related Risks:**

Cryptocurrency-related technologies are not as user-friendly or intuitive for many people, resulting in more user error that results in unwanted outcomes. 'Financial,' 'reputational,' and 'privacy-based' risks are some fundamental categorizations of these user errors. One of the critical points of criticism against the use of Ethereum is the enormous 'gas fees' - the cost of doing business using the technology. Every transaction has many 'gas fees' associated with acting since the Ethereum network is designed to process transactions based on a user performing work and demonstrating proof of work. The details of this concept are beyond the scope of this document.

It needs to be noted that transactions on Ethereum result in users having to shell out substantial sums of money, resulting in a severe critique of Ethereum being just a technology for the already wealthy or the crypto millionaires that made their fortunes when the platform was nascent. Users have sometimes performed erroneous trades or transactions to avoid gas fees while listing/delisting their NFTs. We document such an example below.

### 1.2 Malicious Attacks Such as Phishing:

Phishing or social engineering is often the easiest for malicious actors to perform since it is a low-effort task that is often automated. The attacker can target a broad range of users. The return on investment for this kind of attack is often much higher than the effort required to perform such an attack. These attacks are widespread because the most high-profile owners of NFTs or high-value operators are well-known and easy to study as a target.

Take Todd Kramer, for example, an NFT collector who had acquired Bored Ape and other expensive NFTs, who clicked on a phishing contract that appeared to be a legitimate NFT trader link. Sixteen NFTs from three of his collections were taken, including eight 'Bored Ape' NFTs (a particularly sought-after artwork). In total, the loss totaled around 593 ETH (equivalent to about $1.7 million on February 4, 2022, at 3:55 PM PST). After Todd asked for help on Twitter, OpenSea froze the stolen assets, preventing them from being traded on their platform (White, 2022).

Some commenters noted that the redress (asset freezing and flagging of suspicious accounts) was only possible because OpenSea is a centralized platform with a large amount of power in the NFT arena, which some see as antithetical to ideals of web3. This update also raises the question of whether the creators of the Bored Ape NFT themselves have a way to determine "legitimate ownership" of their NFTs (White, 2022).

**1.3 Scams or Rug Pulls - Where crypto developers accept funds and abscond or cease development activities:**

In the past, there have been instances where customers were a part of exit scams (commonly referred to as rug pulls) specific to NFTs. A rug pull is a malicious maneuver in the cryptocurrency industry where crypto developers abandon a project and run away with investors' funds. ("'Rug Pull' Cryptocurrency Scam Costed Investors over $7.7 Billion in 2021: Chainalysis," 2021) Tron Dogs, a TRX token-based game, created much hype where famous personalities were rooting for it and brought in considerable revenue. TRX is the acronym for Tronix, the in-house cryptocurrency for Tron. Tron is a decentralized digital platform based on blockchain. Tron uses peer-to-peer network technology to eliminate the intermediaries between the content creators and the consumers (Investopedia, 2021).

Tron Dogs did well for six months, but later on, the interaction with the dev team slowed down, and gradually it stopped altogether. In 2020 Tron Dogs was converted to Crop Bytes, thereby bringing relief to the investors (Treak, 2020). Crop Bytes is a crypto farming simulation game in the metaverse where players own a piece of land, raise animals, and perform certain farming activities. Players can also trade in-game items with the other players and earn TRX (Learn, 2021).

Another rug pull example is NiftyMoji, an NFT project where the developers shut down the website's social media and dumped their tokens. NiftyMoji attracted many high-risk investors to invest in their project, which allowed the developers to mine MEXP (NiftyMoji) tokens by selling emojis as NFTs. The developers allegedly pulled off an exit scam by running away with close to a million dollars (Hoogendoorn, 2020). Had there been better security measures in terms of storage of the NFTs, specifically the content, these exit scams could have been avoided.

## Missing or Failing Internal/External Processes:

**2.1 Process Controls:**

Since these platforms are new and moving fast to build a mature economy of NFT trade, there are several gaps or areas of improvement in their product development lifecycle that should mitigate commonly encountered critical bugs and vulnerabilities. We understand that processes like application security review are nebulous concepts still in flux throughout the industry. However, there are some basic processes that these platforms may implement to mitigate at least critical flaws.

Bug bounty programs have been standard industry practice to catch otherwise missed vulnerabilities in the development and testing process. We will now document that OpenSea had an informal bug bounty program that was not that well-known until BleepingComputer and Checkpoint Research documented a vulnerability. This vulnerability showed that a person's NFT balance could be manipulated unauthorizedly. In a report, Check Point researchers summarized the attack as follows:

- Hacker creates and gifts malicious NFT to a target victim.
- The victim views the malicious NFT, triggering a pop-up from OpenSea's storage domain, requesting a connection to the victim's cryptocurrency wallet.
- Victim clicks to connect their wallet and perform the action on the gifted NFT, thus enabling access to the victim's wallet.
- Hackers can get the money in the wallet by triggering an additional pop-up sent from OpenSea's storage domain. The victim is likely to click on the pop-up without reading the note that describes the transaction.

Check Point researchers informed OpenSea of their findings on September 26, 2021. The two parties collaborated to address the issue, and OpenSea came up with a solution less than an hour from the responsible disclosure. OpenSea themselves admit that their bug bounty program was 'informal' until at least October 2021 at the earliest. They have commenced making it formal and publicizing these efforts over the past few months after this disclosure from Checkpoint research.

**2.2 Partnerships: Lack of Industry-wide collaborations (on par with OWASP and MITRe):**

A proactive approach to working with industry stakeholders and participants is vital to mitigate the blind spots in platforms' ability to counter malicious threats and threat actors. Communicating and learning from industry participants and stakeholders has been shown to help industries counter newly emerging threat vectors or behaviors that can result in catastrophic issues.

The MITRe ATT&CK framework is a database of known tactics, techniques, and procedures documented due to a partnership and an understanding of the information security community to share information on how threat actors are orienting their operations. The information on how they are attacking systems is quite helpful in hunting for or detecting such attacks in the wild. Such a partnership in the Web3 community will enable the nimbleness required of companies to mitigate and reduce the volume of attacks on users.

OpenSea claimed to have an 'informal bug bounty program that was not very well-known until they announced an industry-wide security partnership that involves a broad range of technology partners like Adobe, Arweave, and others (Atallah, 2022). Our research group was theorizing an industry-wide partnership. While we were still conducting research, OpenSea announced an NFT security group, an industry-wide partnership to further the security of the NFT web3 ecosystem (Atallah, 2022).

It was unknown to many if OpenSea had a bug bounty program at the time, but they formalized it by bringing in HackerOne to assist with the program in October 2021. In a blog post on their website, they announced their formal bug bounty program on January 12, 2022. Leading up to this announcement, OpenSea had been a highly public target of a few significant security incidents that involved user error and bugs and resulted in some financial loss for a few users.

**2.3 Bugs in Platforms:**

There are many cases in which unforeseen bugs or issues exist with platforms that are poorly documented or understood, even by the platform themselves. Such bugs result in unwanted actions that result in a negative experience for the platform users or may end up in catastrophic losses.

**Example 1:** Glitch kept NFT on sale when users intended to remove them from the sale

Let us take the example of Carson Turner, an NFT collector, who accuses a buyer of exploiting a glitch in OpenSea, and acquiring an NFT that they had no intention of selling.

**Issue:** If a person transfers an NFT listed for sale on OpenSea out of their wallet and back again, it appears not to be for sale despite still being available to buyers. Some people have mistakenly thought they could use this "hack" to delist NFTs if they change their mind about selling them in order to avoid the 'gas fees' (network usage fees) associated with canceling a sale.

This "glitch" resulted in Turner's Bored Ape #2643 NFT being bought even though he thought it was no longer for sale, and he ended up spending 10 ETH (about $38,000) on getting back ownership of this particular NFT.

Twitter user lexomis wrote, "On the human side this kinda is a bummer but it isn't a hack or theft or an exploit. It's being your own bank-level stuff. To be your own bank requires you to understand a lot of these nuances...." (White, 2022)

**Example 2:** Bug in OpenSea platform enables users to purchase NFTs at a massive discount and sell them for a considerable profit

A bug on the non-fungible tokens (NFT) marketplace OpenSea has allowed at least three attackers to secure massive discounts on several NFTs and make a considerable profit.

**Issue**: One attacker, going by the pseudonym "jpegdegenlove," paid a total of $133,000 for seven NFTs – before quickly selling them for $934,000 in ether. Five hours later, this ether was sent through Tornado Cash; a "mixing" service used to prevent blockchain tracing of funds (Gkritsi, 2022). Mixing services are used to obfuscate the financial paper trail, to throw off potential investigations into the provenance of the cash. The concept is sufficient for our conversations, but the technical details are out of the scope of this document.

# External Risks

**3.1 Regulatory Risks:**

Coinbase CEO Brian Armstrong described regulation as one of the significant risks for developing crypto technologies like cryptocurrency and Web3 (Bursztynsky, 2021). In the past decade, cryptocurrencies have gone largely unnoticed by Reserve banks and financial authorities, who took a wait and watch approach to these technologies.

There is a range of opinions on cryptocurrency management through laws and regulations, from the libertarian to the authoritarian. India, for example, outright banned their citizens in dealing with cryptocurrencies, only recently relenting and treating them as assets to be taxed, keeping a ban on their use as financial tender. Reserve banks across the World have made an effort to join the bandwagon of these technologies to craft plans to issue a digital currency. China and India are two nations showing interest in 'digitizing' their national currency, perhaps by using Web3 related technologies, and in tandem, probably restricting the use and development of 'unauthorized tech.'

If regulation stops using current NFT technologies for a significant percentage of people on Earth, it could seriously harm the growth, adoption, and development of web3 and NFT technologies. Ownership verification may not be incorporated into our technology stack and may become an extension of the law rather than an Internet feature. NFT relies on a friendly regulatory regime that allows the ecosystem to thrive and grow without inhibition. That may not always be true of the regulatory environment.

**3.2 Disasters:**

NFTs will have to be protected with appropriate policies related to 'business continuity. Such policies are already critical parts of compliance requirements for companies that want to receive certifications like CIS or NIST 800-53. NFT platforms have to explore the possibility of achieving compliance standards or adopt technology innovations we outline in our solutions section to ensure that external risks are accounted for and a plan exists to mitigate this particular risk.

### 3.3 Geopolitical Risks:

NFTs are newly emerging tech that may become a victim of a type of risk that involves geopolitical conflict or disputes resulting in a significant portion of the World being banned from participating in this technology. When such an event occurs, and it may be hard to predict or foresee, the adoption of this technology will be significantly affected.

## Summary of Risks

To put things in perspective, some of the critical problems with these NFTs platforms are Layer-8 concerns, which entail users unwittingly putting themselves in a vulnerable position that a hacker/attacker can exploit. Another fundamental problem is process/platform related: the storage and ownership of the NFT's content if the developers abandon the project by shutting down the servers, denying collectors access to the NFT. To address these issues, we require a system that ensures that the NFT's content and metadata are permanently stored, allowing users to access their NFTs at any time. In addition, as impersonation is a popular sort of attack, the solution should validate the authenticity of the user/developer associated with a particular NFT.

Given this summary of risks, we propose that what is needed is a censorship-free new decentralized data storage protocol.

## Research on Solutions

We needed to find solutions that could address the issue of NFT ownership and storage while also protecting the platforms from bots. First, we learned about the traditional decentralized storage system, Ethereum, which has its own set of drawbacks. Ethereum was initially designed to be around 500GB - 1TB in size, which is not ideal for large amounts of data, and the cost of deploying this much data would be prohibitively expensive due to the high gas fees. Because of these constraints, Ethereum is not, by itself, an adequate solution. In order to move away from Ethereum and find a viable solution, specific characteristics - Persistence Mechanism, Data Retention Policy, Decentralization, and Consensus were taken into consideration (Richards, 2021).

Persistence mechanisms are used to ensure that data is stored and accessible indefinitely. Persistence mechanisms are classified into two types: blockchain-based and contract-based. Blockchain-based persistence accounts for the entire chain to replicate data across all nodes. Contract-based persistence saves the hash of the data's location. The data stored on this chain is not replicated but rather renewed over time. Because we focus on reducing scams and rug pulls, blockchain-based persistence would be ideal for data preservation. The Interplanetary File System (IPFS) is a traditional method of ensuring long-term persistence by pinning data to the network. Arweave, in addition to Ethereum, is a blockchain-based persistence platform that will be discussed further below (Richards, 2021).

Data retention is another important consideration when looking for a solution to ensure the storage of NFT metadata. This goal can be accomplished by utilizing some cryptographic challenge issued to the nodes to verify that the data is still stored. Arweave ensures that all of these boxes are checked by utilizing a "Proof-of-Access" consensus mechanism and storing the data indefinitely (Richards, 2021). A *consensus mechanism* is a system that prevents certain types of economic attacks and enables distributed systems to collaborate while remaining secure.

It was crucial to find a solution to the bots exploiting NFT platforms that would produce the least false positives while also reducing the risk of unwanted automated account takeover attacks. Moreover, we wanted a solution which would flag the use of compromised credentials to protect against credential stuffing attacks. Because of its large and trusted customer base, PerimeterX piqued our interest and prompted us to investigate further.

# Interplanetary File System - IPFS

Small and new NFT marketplaces use centralized servers. They have been trying to move to a decentralized technology such as the Interplanetary File System (IPFS), a protocol and a peer-to-peer network that allows data to be stored and shared in a distributed file system. IPFS addresses a file by its content and identifies it by its location. The content identifier in IPFS is a cryptographic hash of the content at the address mentioned ("/ipfs/"). Since the address of the file is created from the file's content, the links in IPFS cannot be changed. Content identifiers use hash functions to check for file integrity and enable users to verify that a file has not been modified since it was published (IPFS, n.d).

Although IPFS may seem a possible solution, it does come with its fair share of limitations. One, file sharing in IPFS is only possible by sharing the file link through traditional communication channels, i.e., messaging, email and other social media messaging platforms. Sharing these links via such platforms can increase the likelihood of layer-8 risks, leading to increased volumes of malicious links being circulated. Second, file discovery can take much time and cause a bad user experience, harming the marketplaces' business reputation and causing nodes to lose data. There are centralized services like Pinata that can help pin these files but integrating two services will be complex and expensive to the small marketplaces that do not have an advanced tech team (Geier, 2020).

# Arweave

Arweave is an information security company that has created a censorship-free new data storage protocol on a blockchain-like structure called the blockweave. A Blockweave is very similar to the standard blockchain technology with just one difference: blockweave performs distinct block verification on the Arweave Permaweb. This decentralized storage network permanently stores data on its chain. Unlike Bitcoin, where miners are asked to solve a complex mathematical problem, a process called - proof of work, Arweave requires computers to verify that a new set of transactions contain a randomly selected marker from previous transactions on the network. Only if the marker is correct can the transactions be added to the network, and by doing so, the computer that adds the transaction is rewarded in the form of Arweave (AR Cryptocurrency); this process is known as "Proof-of-Access" (Advani, 2018).

In order to ensure that decentralization is secure, Arweave introduced a new concept called 'Blockshadows.' Traditionally, an entire block is broadcasted during a transaction. In contrast, Arweave transmits a shadow that is then reconstructed by the recipient nodes. The blockshadow only holds the list of transaction-hashes rather than the transactions themselves. Arweave uses less energy than traditional solutions by decoupling these transactions and creating a low-cost network (Advani, 2018).

One primary security concern with NFTs is authentication problems, which has caused many NFT holders to lose money, where hackers have sold NFTs by claiming to be the owner of them (Sinclair, Kim 2021). Although authentication is a security concern, the problem arises when users are caught off-guard by hackers. Arweave solves this problem by recording the author's identity who commits a file to the blockchain, which is readily accessible throughout the file's existence. This recording mechanism helps the owner prove their identity and certify its authenticity to anyone accessing the file or at any time a transaction is being made, thereby protecting against the Layer-8 authenticity risks (Advani, 2018).

The token of the NFT is decentralized and is stored on the blockchain. These tokens point to the HTTP URLs, which are centralized in nature and can never be truly "ownable" since they depend on the organization always to have the server operating (Kahan, 2021). However, the storage for the metadata or the content is decided by the developer minting it. The security/storage of the metadata solely depends on the companies maintaining the centralized servers.

One of the examples where the NFT is hosted on the marketplace with elements of centralized storage is Beeple's "Crossroad" on Nifty Gateway. This NFT's token directs to the HTTP URL hosted on Nifty Gateway, which contains the metadata. The metadata text points then direct to another HTTP URL containing the visual media hosted on a cloud-based service served by the Nifty Gateway's servers. Owning this kind of NFT can be risky since the owner/collector can lose access to the NFT and just be left with a token referring to a dead HTTP URL if Nifty Gateway was to discontinue their operations (Kahan, 2021).

1111 by Kevin Abosch, InfiNFT Minting Platform, and Mintbase are a few examples of the NFT being hosted on a marketplace with the decentralized storage of metadata and visual assets. 1111 is a piece of art by Kevin Abosch that was released as an NFT on OpenSea in March 2021. This NFT's metadata and photos are saved on the Arweave Permaweb. NFT42's InfiNFT is a new NFT minting platform that employs Arweave to store the art completely on-chain. It focuses on token integrity and ensuring that the creations are available indefinitely. Mintbase.io is a platform that enables individuals and organizations to create NFT tokens for various uses, including artwork and tokens for products and services. In order to tackle the issue of centralized storage and faster data availability, they have integrated Arweave for their backend storage (The Arweave Project, 2020).

Comparing the two examples above and looking at them from an investor/collector's perspective, it is evident that one would want to own an NFT with the content and metadata stored in a decentralized network. From a technical point of view, most NFTs have been built on the Ethereum network, which has been facing many issues about its high gas fees because of the high volume and demand of NFT transactions (Benoliel, 2021). No NFT marketplace would want its customers to face the potential threats that centralized storage poses.

## Arweave and Partners

Solana was officially launched in March 2020 to encourage the creation of smart contracts and decentralized applications by implementing a high-performance and high-frequency blockchain. The blockchain operates on two concepts, "Proof-of-History" (PoH) and "Proof-of-Stake" (PoS). PoH is derived from PoS, where PoS allows computers on the network to verify the transactions depending on the number of coins it holds. PoH allows these transactions to be time-stamped and verified faster (Locke, 2021).

In order to create less data congestion on the blockchain, Solana partnered with Arweave to announce the SOLAR Bridge. This bridge allows Solana to move and store its transaction history to a dedicated blockchain or storage network. This partnership allows Solana to solve its crucial issue, data storage. Solana has a higher demand for data storage since all the historical data needs to be stored forever. By outsourcing the storage to Arweave, Solana has helped Arweave gain more users on the network and open doors for other blockchain protocols to partner up with Arweave (Kong, 2020).

Polygon's partnership with Arweave is another example of a blockchain network outsourcing data storage to Arweave. Polygon, previously known as Matic Network, was founded to solve Ethereum's scalability problem. It is a platform designed to launch interoperable blockchains. In 2020, it attracted much attention from developers and investors since the transaction cost on Ethereum was constantly increasing. Polygon is a four-layer system - Ethereum layer, Security layer, Polygon Networks layer, and Execution layer. The Ethereum layer consists of a set of smart contracts implemented on Ethereum. The security layer allows blockchains to leverage the benefits of this layer which provides additional security. The Polygon networks layer is the ecosystem of the blockchain networks built on Polygon. Fourth is the execution layer responsible for executing the smart contracts through Polygon's Ethereum Virtual Machine (Phillips, 2021).

This partnership allows Polygon users to pay for data uploads using Polygon's MATIC token and storing smart contracts on the permaweb. Arweave's integration with Polygon through Arweave's Bundlr network enables developers to make decentralized storage available inside Polygon's applications. Moving on, Polygon aims to make this decentralized storage available directly to the end-users by integrating this capability in all the wallets in the Polygon ecosystem. To build this integration into the wallet infrastructure for the end-users, the three parties - Polygon, Arweave, and Bundlr Network are offering a $9000 grant to the developers (Polygon Team, 2021).

By integrating Arweave in their systems, small marketplaces will scale their business by handling large volumes of data that blockchain technologies cannot. In order to tackle the issues of storage and ownership of NFTs, major marketplaces like OpenSea or Nifty Gateway should partner with Arweave, similar to what Solana and Polygon have done for data storage on the permaweb.

## PerimeterX

Keeping automated bots from impacting NFT sales and artificially inflating NFT prices is critical. Artificially inflating the prices of NFTs or making participation in limited NFT drops or launches are a form of [1.2 malicious attacks] risk-type. Bots can be mitigated, deterred, blocked, and descented, although it is nearly impossible to do so. PerimeterX is the leading digital enterprise security solutions provider that protects against automated fraud and client-side threats. It contains a bot detection and blocking tool that uses machine learning, behavioral analysis, and prediction algorithms to detect and stop bots (PerimeterX, 2022). PerimeterX appears to be the appropriate solution for bridging the gap between Web2 and Web3 settings since it can be smoothly incorporated into both systems with the existing Contact Delivery Network (CDN).

Bot Defender by PerimeterX is a behavioral bot management solution protecting websites and mobile apps from automated bot attacks. It employs a user-friendly verification method that protects against CAPTCHA-solving bots and enhances the user experience. Detect, Enforce, and Report are the three fundamental components of the Bot Defender. It reduces user friction by combining intelligent fingerprinting and behavioral signals to detect bots on the web and API endpoints. The Bot Defender offers a variety of enforcement actions across all client contexts to reduce the impact of bots on the marketplace (PerimeterX, n.d).

Although NFT marketplaces do not currently have PerimeterX integrated into their environment, the Bot Defender is integrated into the environments of other large corporate companies. Crunchbase, the leading source of the business information on private and public companies, is a PerimeterX customer dealing with the problem of legitimate users being flagged as bots. PerimeterX provided a deployment option that integrated with Crunchbase's existing infrastructure and provided them with flexible control over traffic filtering. Crunchbase reduced false positives and improved user experience, and authorized users can now access critical resources when needed by integrating the Bot Defender into their environment (PerimeterX, n.d).

# Recommended Best Practices

## Defend Against Bots

Implementing the PerimeterX Bot Defender is a multi-step process that involves configuring the PerimeterX environment, deploying and tuning the Bot Defender, and finally mitigating the bots. A first step towards integrating this environment in the marketplace is to set up PerimeterX Sensor and PerimeterX Enforcer. Deploying these would allow the Sensor, cloud-based detector, and Enforcer to integrate a wide range of server-side technologies, which would allow the marketplaces to preserve their existing architecture. In order to protect against bots and other attacks such as brute force, it is essential to tune the settings in the environment. This process leverages the machine learning algorithms according to the rules set by the client regarding the type of attacks - credential stuffing, account takeover, bots [2.3 risk type].

PerimeterX identifies the office space, known partners, and other tools used for developmental purposes to ensure that the incoming traffic on the client-side is not affected by these algorithms. Additionally, the Bot Defender also uses a threat intelligence framework that reduces false positives, increases the number of legitimate users, and avoids pitfalls of CAPTCHA, thereby ensuring maximum security from bots. Finally, with all these frameworks set up in the marketplace environment, the Bot

Defender would create dashboards for the client to assess the traffic/bots on their network and accordingly help in mitigating it.

Stopping automated web attacks, including the most advanced account takeover attacks, is also part of defending against bots. Until recently, financial services companies were the primary target of account takeover attacks, but the focus has shifted to cryptocurrency platforms and NFT marketplaces. The Bot Defender employs behavior-based and predictive analytics to detect modern account takeover attacks and identify patterns in traffic. Most importantly, it is not configured to use the existing rules to distinguish bots from real users and detect bogus account creation attempts in real-time.

## Solve for Permanence and Ownership

Taking inspiration from the above-mentioned Arweave partnerships, marketplaces that support on-chain media storage solutions such as Arweave should leverage those benefits. Those that do not support an on-chain media storage solution should adopt a solution to ensure the storage and ownership of the NFTs. Marketplaces like OpenSea support storing the metadata on decentralized networks such as Arweave. OpenSea allows the deployment of the metadata API according to the user's storage needs, for which OpenSea has created sample API servers in Python and NodeJS. This feature allows developers to store the metadata and content of the NFT on Arweave and eliminate the risk of customers getting their crypto wallets wiped off in case of scams or rug pulls [1.3 risk type]. By hosting NFTs on Arweave, the URL appears in a different format - "ar://<hash>" which can educate the collectors/buyers to collect Arweave hosted NFTs, thereby reducing the effects of Layer-8 risks (OpenSea, n.d).

The Arweave blockchain is used to power ArDrive, a persistent data storage solution. It develops technology that enables digital data to be permanently stored. When you use ArDrive, all of your files become part of the Arweave blockchain, and you can choose to keep them private or share them. ArDrive is cost-effective for customers since it avoids the subscription model, where users are forced to pay for storage that they do not require. It operates on a Pay-Once model, which means that the user only pays once for any file they upload (ArDrive, 2022).

NFT marketplaces and ArDrive are a natural fit because the former allows users to buy, sell, and transfer ownership of NFTs. At the same time, the latter lets users save the image and metadata of NFTs indefinitely [2.2 risk type]. Marketplaces currently use the Interplanetary File System to freeze metadata, ensuring that it cannot be modified but not permanently stored. ArDrive saves copies of data on the blockchain and stores and distributes data across numerous hosts. Furthermore, any publicly shared data is encrypted independently using a unique hash, adding an extra degree of security to the data's protection from intruders.

Although IPFS appears to be a viable solution for marketplaces to adopt, it can only store one copy of data. This characteristic may not be a problem in the short term, but backing up large amounts of data would necessitate multiple copies on the nodes in the long run. If all marketplaces choose IPFS as their solution, Arweave has released an Arweave+IPFS bridge that allows the content of the IPFS file to be permanently stored. This bridge has been tested on several nodes for more than six months, and the results have been better than expected, so it has been made available to the public (The Arweave Project, 2019). Marketplaces that do not want to switch to a different blockchain can undoubtedly use this bridge to ensure the longevity and ownership of NFTs.

One key factor to consider is the current architecture, in which NFTs are minted on Ethereum, which causes severe congestion on the server due to increased network traffic. This problem can be remedied by switching to Arweave, which uses a load-balancing technique that allows hosts to cache frequently requested data without calling the servers repeatedly. This will reduce server load, reduce network traffic, and allow for faster data transmissions. As data is kept on multiple nodes, Arweave's Peer to Peer technology assures that file transfers and downloads are faster by removing the risk of network bottlenecks.

## Summary of Solutions

- **NFT platforms should implement processes and systems that prevent unauthorized transactions of NFTs**

Such actions should mitigate layer-8 risks, particularly malicious attacks and phishing risks [1.2 malicious attacks]. These mitigations can take the form of implementing Arweave as a possible data storage protocol

- **NFT platforms should store metadata on the blockchain**

Architecting a blockchain-based storage solution automatically mitigates certain risks in layer-8 and offers data permanence. Such actions ensure that disasters and geopolitical risks [3.2 and 3.3 risk types] are also drastically reduced while ensuring that NFT collectors still have access to the metadata in case of rug pulls, a form of [1.3 scams or rug pulls] risk-type

- **NFT platforms should clearly and organically record the identity of the owner of the NFT and offer easy verification**

Leveraging a blockchain solution other than Ethereum that records the owner's identity would ensure that NFTs are traded ethically, and users do not fall for malicious

NFTs. Moreover, with a lack of regulations, such a solution would ensure that ownership is included in the technology stack [3.1 Regulatory Risks]. Without proper security measures, malicious attackers can quickly get past the user authentication and access the NFT's content [2.1 and 2.2 risk types]

- **NFT platforms should focus on improving scaling by working on reducing user costs, such as gas fees**

Shifting to newer technologies would reduce network congestion by recording hashes and not blockchain transactions. Such actions would reduce financial risks

- **NFT platforms should implement user training about the platform**

User errors in cybersecurity are very difficult to mitigate. By implementing a user training that trains the users about two-factor authentication and malicious links can reduce user-related privacy risks [1.1 Inadvertent or User Error]

- **NFT platforms should integrate a bot defending environment to protect against bots**

By integrating a bot defending software in their environment, NFT platforms can successfully protect against bots. Moreover, this would allow the platforms to mitigate account takeover attacks and prevent attackers from artificially hiking the NFT prices, a form of [2.3 bugs in platforms] risk-type

# Acknowledgements

This paper and the research behind it would not have been possible without the consistent and exceptional support of Professor Andrew Reifers at the University of Washington Information School. His eagerness, knowledge, and attention to detail have been invaluable in keeping this paper on track from the first time we discussed the research scope to the final draft. I'm also grateful to Jayanth Kumar, a security analyst at Kalles Group. His understanding of identifying risks associated with NFT platforms was extremely helpful in categorizing the risks and working on identifying viable solutions. Professor Reifers' and Jayanth's expertise have greatly enhanced this study.

# Citations